\documentclass[aps,prl,prabib,twocolumn,showpacs,nofootinbib]{revtex4}
\usepackage{graphicx} \usepackage{amsmath} \usepackage{amssymb}
\usepackage{amsfonts} \usepackage{bm}

\begin{document}

\newcommand{\be}{\begin{equation}} \newcommand{\ee}{\end{equation}}
\newcommand{\bea}{\begin{eqnarray}}\newcommand{\eea}{\end{eqnarray}}

\title{Quantization of neutron in Earth's gravity}

\author{Pulak Ranjan Giri} \email{pulakranjan.giri@saha.ac.in}

\affiliation{Theory Division, Saha Institute of Nuclear Physics,
1/AF Bidhannagar, Calcutta 700064, India}

\begin{abstract}
Gravity is the weakest of all four known forces in the universe.
Quantum states of an elementary particle due to such a weak field is
certainly very shallow and would therefore be an experimental
challenge to detect. Recently an experimental  attempt was made by
V. V. Nesvizhevsky et al., Nature ${\bf 415}$, 297 (2002), to
measure the quantum states of a neutron, which shows that ground
state and few excited  states are $\sim 10^{-12}$ eV. We
show that the energy of the ground state of a neutron confined above
Earth's surface should be $\sim 10^{-37}$eV. The
experimentally observed energy levels are $10^{25}$ times deeper
than the actual energy levels it should be and thus certainly not
due to gravitational effect of Earth. Therefore the correct
interpretation for the painstaking experimental results of Ref.
\cite{nes1} is due to the confinement potential of a one dimensional
box of length $L \sim 50\mu$m, generated from the experimental setup
as commented before \cite{hansoon}. Our results thus creates a new
challenge to the experimentalist to resolve the shallow energy
levels of the neutron in Earth's gravitational field in future.

\end{abstract}

\vspace{1cm}


\date{\today}

\maketitle

The investigation of quantum phenomenon in gravitational field is
certainly interesting and challenging \cite{nes1,nes3,peters,ber} due
its weakness of strength. To get an idea of the weakness of
gravitational force over other forces a quantitative estimation may
be helpful. The the gravitational attraction of two neutrons
separated by a distance $r$ is $\sim 10^{-36}$ times weaker
\cite{hartle} than the Coulomb repulsion between  the two electrons
separated by the same distance. One therefore needs to be very
careful while investigating the quantum effects of gravity. Neutron
is a possible candidate on which quantum effects of gravity can be
investigated because  charge neutrality will eliminate
electromagnetic force from our considerations.

The nature of the gravitational force $F$ of Earth (except the
strength) experienced by a neutron is same (long range and
proportional to the inverse of the distance between the two agents)
as that of the Coulomb force experienced by an electron in a
Hydrogen atom. It is therefore expected that the nature of energy
states of a neutron in the Earth's gravitational force will be
similar to that of a  Hydrogen atom with an infinite hard sphere
core \cite{care,meyer}. We need to keep in mind that the neutron is
above the Earth's surface, so we assume that the wave-function
within the Earth is zero, i.e.,  $\psi(r)=0$ for $r\leq R_{\oplus}$,
where $R_{\oplus}$ is the Earth's radius (it is assumed that Earth
is completely spherical). Since the neutron of mass $m$ can not
penetrate within the Earth, it will put an upper bound to the
absolute value of the energy $E_n$  of the discrete quantum states,
which is $|E_n|\leq (\hbar^2/2m)R_{\oplus}^{-2}\approx 5.08\times
10^{-37}$eV \cite{care,meyer}. Note that the states are
$\sim10^{-25}$ times less deeper than that obtained in the recent
experiment by V. V. Nesvizhevsky et al., Nature ${\bf 415}$, 297
(2002).

Then the question arises that what is the reason of getting quantum
states $\sim 10^{-12}$eV in the experiment? The correct
interpretation for observing $\sim$peV  ($1$peV= $10^{-12}$eV)
states  in the experiment is the following. The experimental set up
consists of a bottom mirror and a top absorber with a gap of
approximate $50\mu$m in between them. This can be considered as a
problem of a particle in a one dimensional box \cite{landau} of
length $L= 50\mu$m. In fact it has  been commented before in Ref.
\cite{hansoon}, see the corresponding reply \cite{nes4} also. For
the present purpose we may neglect the dynamics of the neutron in
the transverse direction. The energy levels for the neutron in the
potential created by the box is $E_n= -(\hbar^2\pi^2n^2/2m)L^{-2}$.
The first few states for $L= 50\mu$m are respectively given by
$E_1\approx 0.082$peV, $E_2\approx 0.3272$peV, $E_3\approx
0.7362$peV, $E_4\approx 1.309$peV, $E_5\approx 2.0451$peV, $E_6
\approx2.945$peV, $E_7\approx 4.01$peV. Note that  the
experimentally obtained first four energy levels in Ref. \cite{nes1}
are comparable with the above obtained theoretical levels $E_4$,
$E_5$, $E_6$ and $E_7$ respectively.

We then need to answer what is wrong with the previous theoretical
prediction of Ref. \cite{nes2}, which shows that  the discrete
quantum levels due to Earth's gravitational force is $\sim$ peV?
In fact it agrees with the experimental results \cite{nes1}. The
answer could be found partly in the potential $U(z)= mgz$ considered
for the neutron above the Earth's surface. The other drawback is
that the spherical symmetry of the problem due to central force has
been completely ignored and  thus the dynamics of the neutron in the
$z$ direction has be decoupled by assuming that in the transverse
directions the particle is  free. It is of course true that the
potential $U(z)$ is approximately valid for $z\ll R_{\oplus}$. But
for the neutron above the Earth  the wave-function  in principle
would extend from earth's surface to infinity. Thus $U(z)$ is
useless in this situation and instead should be replaced by
spherically symmetric  Newton's potential $V(r)= -GM_{\oplus}m/r$
\cite{penrose}, where $G$ is the universal gravitational constant,
$M_{\oplus}$ is the mass of the Earth and $r$ is the distance of the
neutron from the Earth's center. The potential within the Earth is
as usual infinity because of the assumption that the probability of
finding the neutron within the Earth is zero. The ground state of
the neutron on the Earth's surface in gravitational field is thus
$E_{\mbox{g.s}}\approx-(\hbar^2/2m)R_{\oplus}^{-2}\approx -5.08\times
10^{-37}$eV \cite{care,meyer}, since it is the maximum deep state of the
neutron. The analytical calculation for all
the excited states of the neutron will be in line with
Ref. \cite{meyer}. However 
analytical  solutions for excited states are not important for our
present purpose because they all will be even less than the ground
state energy. The point that the deepest bound state (which is
ground state) is $\sim 10^{-37}$eV is the most important message
here. However one needs to think about the validity of the basic
assumption in reality  that the neutron wave-function within the
Earth is zero. Because, the penetration of the quantum particle
probability within the Earth will change the bound on the quantum
energy levels of the neutron. But this is an issue which can be best
resolved by experimental observation. We however have considered
this assumption based on the experiment \cite{nes1} The next
immediate challenge to the experimentalist is to detect the quantum
states due to Earth's gravity.

Our theoretical observation does not rule out the experimental detection of
$\sim 10^{-12}$eV quantum states of a neutron but rather it gives a correct
interpretation for the existence of peV states. The one dimensional box
potential generated from the lower mirror and top absorber of length
$\sim 50\mu$m  dominates the
gravitational potential on the Earth's surface. Gravitational force on the
Earth is so weak that the quantum states of a neutron due to such force is 
$\sim 10^{-37}$eV, based on the assumption  that neutron
wave-function inside the
Earth is zero. The experimental resolution power should be much higher than
the present we have \cite{nes1} in order to detect such quantum  states.




\end{document}